\documentclass[preprint,prc,showpacs,preprintnumbers,amsmath,amssymb,floatfix]{revtex4-2}
\usepackage{amsmath}
\usepackage{graphicx}
\usepackage{dcolumn}
\usepackage{bm}
\usepackage{ulem} 
\usepackage[usenames]{color}
\usepackage{epstopdf}
\usepackage{epsfig}
\usepackage{float}
\usepackage{subfigure}
\usepackage{multirow}
\usepackage[colorlinks,linkcolor=blue,anchorcolor=teal,citecolor=red]{hyperref}

\newcolumntype{R}{>{\raggedleft\arraybackslash}X}

\newcommand{\nc}{\newcommand}       
\nc{\vc}[1] {\mbox{\boldmath $#1$}} 
\nc{\del}       {\partial}              
\nc{\bra}       {\langle}               
\nc{\ket}       {\rangle}               
\nc{\bras}[1]   {\langle #1|}           
\nc{\kets}[1]   {|#1\rangle}            
\nc{\mapleft}[1]{           
	\smash{\mathop{\,          %
			\hbox to 1.5cm{\rightarrowfill}\, }\limits_{#1}}}
\nc{\beq}     {\begin{eqnarray}} \nc{\eeq}    {\end{eqnarray}}
\nc{\nn}      {\\\nonumber} \nc{\vs}      {\vspace{-0.275cm}}
\nc{\fra}    {\frac{1}{2}}
\nc{\mb}        {\mathbf}

\begin{document}

\preprint{}

\title{The comparison of the state-of-the-art nucleon-nucleon potentials from phase shift to nuclear matter}

\author{Ke Nan}
\affiliation{School of Physics, Nankai University, Tianjin 300071,  China}
\author{Jinniu Hu}
\email{hujinniu@nankai.edu.cn}
\affiliation{School of Physics, Nankai University, Tianjin 300071,  China}
\affiliation{Shenzhen Research Institute of Nankai University, Shenzhen 518083, China}
\author{Ying Zhang}
\affiliation{Department of Physics, Faculty of Science, Tianjin University,Tianjin 300072, China}
\author{Hong Shen}
\affiliation{School of Physics, Nankai University, Tianjin 300071,  China}
\date{\today} 

\begin{abstract}
The nucleon-nucleon ($NN$) potential is the residual interaction of the strong interaction in the low-energy region and is also the fundamental input to the study of atomic nuclei. Based on the non-perturbative properties of the quantum chromodynamics (QCD), $NN$ potential is not yet directly accessible from QCD theory. Therefore, various models of $NN$ interactions have been constructed based on Yukawa's meson exchange pictures since the 1930s, including one-boson-exchange models, coordinate operator models and chiral effective field models. Analysis of extensive $NN$ scattering data has shown that the two-body nuclear force exhibits a short-range repulsion and intermediate-range attraction, and decays rapidly with increasing distance. A series of charge-dependent high-precision $NN$ interactions have been further developed in the past thirty years, such as the AV18 potential, CD-Bonn potential, pvCD-Bonn potentials, and the chiral effective nuclear potentials with momentum expansion up to the fifth order. In this work, the phase shifts at different channels, the cross sections, the entanglement entropy in spin space, and the equations of state of symmetric nuclear matter and pure neutron matter from these high-precision $NN$ interactions are calculated and systematically compared. It can be found that they have significant differences in the cases with high angular momentum, high laboratory energy, and high-density regions. 

\end{abstract}


\maketitle

\section{Introduction}

The nucleon-nucleon ($NN$) interaction is a fundamental aspect of nuclear physics, governing the intricate quantum behavior of protons and neutrons within atomic nuclei. An effective $NN$ force must precisely represent various characteristics, including strong repulsion at short distances, strong attraction at intermediate ranges, and diminishing to zero at long distances. Additionally, it must conform to fundamental symmetries including spatial rotation invariance, translational invariance, and space reflection invariance \cite{epelbaum09,machleidt11}.

The $NN$ interaction represents the residual effect of the strong interaction at the level of nucleons, {which is governed by quantum chromodynamics (QCD) theory}. Early models, like Yukawa's one-pion-exchange model, laid foundational groundwork \cite{yukawa35}. In the 1960s, the introduction of heavy bosons aimed to resolve short-range and intermediate-range $NN$ interaction challenges \cite{erkelenz74}. Subsequently, the Bonn University group developed a comprehensive meson exchange model, culminating in 1987 with its final version \cite{machleidt87,machleidt89}.

By the 1990s, meticulous analysis of extensive $NN$ scattering data revealed significant differences in proton-proton ($pp$) and neutron-proton ($np$) scattering phase shifts after removing the Coulomb interaction contributions,  prompting further investigations into charge dependence in nuclear interactions \cite{stoks94}.

Charge independence in isospin space remains invariant under any rotation, while violations of this symmetry constitute charge-dependent or charge independence breaking (CIB). Charge symmetry breaking (CSB) specifically refers to deviations under a 180° rotation about the $y$-axis in isospin space, where the positive $z$ direction correlates to positive charge. Thus,  CSB, represents a subset of charge dependence, where strong $NN$ interaction CIB indicates differences in interactions among $pp$, $np$, and neutron-neutron ($nn$) pairs in the isospin $T=1$ state, after eliminating electromagnetic influences.

Since then, several high-precision $NN$ interaction models have been established, such as Reid93, Nijmegen93, Nijmegen I, Nijmegen II \cite{stoks94}, AV18 \cite{wiringa95}, and CD-Bonn \cite{machleidt01}. Additionally, the chiral $NN$ potential, based on chiral perturbation theory, has played a pivotal role \cite{weinberg90,weinberg91,weinberg92, ordonez94,ordonez96,epelbaum98,epelbaum00,entem03,epelbaum05,entem15,epelbaum15a,epelbaum15b,entem17,reinert17}. {The relativistic versions of chiral potentials, developed by the Beihang group, extend from leading order to next-to-next-to-leading order~\cite{ren17,xiao19,lu22}. These advancements provide a more accurate description of NN scattering phase shifts at equivalent chiral expansion orders, as well as the saturation properties of symmetric nuclear matter ~\cite{zou24}.}  Recent contributions from the Granada group have explored phase shifts with delta potentials, addressing associated uncertainties \cite{perez14,arriola20}. While these potentials effectively reproduce $NN$ scattering data, variations in phase shifts under high partial waves and momentum regimes warrant detailed comparison.  

{Over the past 20 years, significant advances have been achieved in understanding baryon-baryon interactions, particularly $NN$ potentials, through lattice QCD simulations. These efforts have successfully captured many of the fundamental characteristics of NN potentials ~\cite{ishii07,inoue11,beane13,mcllroy18,lyu21,lyu23}. Moreover, recent work has aimed to connect earlier model-dependent NN potentials with results obtained from lattice QCD theory~\cite{hu16,song18,bai20,hu20}.} 

The scattering amplitude, which governs cross sections, differential cross sections, and spin polarization observables in $NN$ scattering, are encapsulated by the $S$-matrix through partial wave expansion. This unitary $S$-matrix parameterization, reliant on phase shifts and mixing angles across angular momentum channels, accurately describes these observables \cite{arriola20}, with most high-precision $NN$ potentials achieving $\chi^2/N \sim 1$. Furthermore, detailed comparisons of cross sections across various input nucleon energies remain pertinent.

Recent discussions on $NN$ scattering entanglement in spin space and Wigner symmetry within the framework of chiral effective field theory have invoked concepts from quantum information theory. For instance, $np$ scattering entanglement potency in the $S$-channel correlates with phase shift distinctions at $^1S_0$ and $^3S_1$ channels \cite{harshman05,beane19,low21,bai23,hu24}. $NN$ potentials have also facilitated {\it{ab initio}} investigations into finite nuclei and infinite nuclear matter using methods such as the variational approach, the Brueckner-Hartree-Fock method, and the relativistic Brueckner-Hartree-Fock method \cite{caia02,lacombe75,jackson75,drechsel92,drechsel99,fuchs98,brockmann90,hu17,shen19,wang22}.

This study compares 17 high-precision $NN$ potentials encompassing AV18 \cite{wiringa95}, CD-Bonn \cite{machleidt01}, pvCD-Bonn \cite{wang19}, and chiral effective field \cite{epelbaum15a,entem17,reinert17}. With these potentials, we analyze the phase shifts, cross sections, entanglement potency, and nuclear matter equations of state across different laboratory energies and densities. This paper is structured as follows: Section \ref{sec2} presents phase shift comparisons across the 17 $NN$ potentials, Section \ref{sec3} showcases their differential cross sections, Section \ref{sec4} details entanglement power calculations, and Section \ref{sec5} explores equations of state for nuclear matter. Finally, Section \ref{sec6} concludes with a summary and implications.

\section{Nucleon-nucleon phase shifts}	\label{sec2}
The Lippmann-Schwinger equation can be directly applied to study $NN$ scattering processes, which are equivalent to considering the scattering boundary conditions in the Schr\"{o}dinger equation~\cite{machleidt01}, 
\begin{equation}
	\left[\frac{d^2}{dr^2}+q^2-\frac{l(l+1)}{r^2}-MV\right]\chi_{l}(r;q)=0,
\end{equation}
where, $l$ is the orbital angular momentum, $V$ is the $NN$ interaction, and $q$ is the momentum of the input particle. The asymptotic behavior of the radial wave function $\chi_l$ determines the scattering phase shift. If $V$ is not a long-range force, the asymptotic wave function of $\chi_l$ is the Riccati-Bessel function.

The Lippmann-Schwinger equation introduces the $T$ matrix, which can be expressed as an integral equation related to the interaction, 
\begin{equation}
	T(\vec{p'},\vec{p})=V(\vec{p'},\vec{p})+\int d^3\vec{k}V(\vec{p'},\vec{k})\frac{1}{E-E_{K}+i\epsilon}T(\vec{k},\vec{p}).
\end{equation}
For computational convenience, we usually solve the Lippmann-Schwinger equation in the partial wave representation. In this representation, the $T$ matrix and the interaction potential $V$ can be expressed as,
\begin{eqnarray}
	&&T_{ll'}^{sj}(p',p)=\langle p',l',s,j|T|p,l,s,j\rangle,\\ \nonumber
	&&V_{ll'}^{sj}(p',p)=\langle p',l',s,j|V|p,l,s,j\rangle.
\end{eqnarray}
In $NN$ scattering, the total angular momentum and total spin are conserved. Therefore, in the partial wave representation, the Lippmann-Schwinger equation can be formulated as,
\begin{align}
	T_{ll'}^{sj}(p',p)=V_{ll'}^{sj}(p',p)+\sum_{l''}\frac{2}{\pi}\int_{0}^{\infty}dkk^2V_{ll''}^{sj}(p',k)\frac{1}{E-E_{k}+i\epsilon}T_{l''l'}^{sj}(k,p).
\end{align}

There are various methods for numerically solving the Lippmann-Schwinger equation, with one common approach being the matrix inversion method proposed by Haftel and Tabakin in 1970~\cite{haftel70}. This method employs Gaussian integration to transform the integral into a numerical summation, thereby converting the Lippmann-Schwinger equation into a matrix form,
\begin{eqnarray}\label{LS}
	T_{ll'}^{sj}(p',p)=V^{sj}_{ll'}(p',p)+\frac{2}{\pi}\sum_{l''}\sum_{i=1}^{N_p}w_{i}k_{i}^{2}V_{ll''}^{sj}(p',k_{i})\frac{2m}{p^2-k^{2}_{i}+i\epsilon}T_{l''l'}^{sj}(k_{i},p),
\end{eqnarray}
where $k_{i}$ and $w_{i}$ represent momentum grid points and integration weights, respectively. In practical computations, Gauss-Legendre integration is frequently used. $N_{p}$ is the number of momentum grid points, and with a sufficiently large $N_{p}$, numerical integration achieves high accuracy. By moving the part of Eq. (\ref{LS}) related to $T$ to the left side, the matrix equation $\mathbf{(1-VG)T=V}$ is obtained, and the $T$ matrix is ultimately determined by matrix inversion: $\mathbf{T=V(1-VG)^{-1}}$. However, it's worth noting that since the propagator $G=\frac{2m}{p^2-k^{2}_{i}+i\epsilon}$ contains complex terms, the resulting $T$ matrix should be represented as complex numbers. To facilitate practical computation, a purely real $R$ matrix is introduced. By defining the $R$ matrix as,
\begin{equation}
	R=T+i\pi T\delta(E+H_{0})R,
\end{equation}
the Lippmann-Schwinger equation can be expressed in a purely real form,
\begin{equation}
	R(\vec{q'},\vec{q})=V(\vec{q'},\vec{q})+P\int d^{3}\vec{k}V(\vec{q'},\vec{k})\frac{M}{q^{2}-k^2}R(\vec{k},\vec{q}).
\end{equation}
This equation can be solved in partial wave representation similar to computing the $T$ matrix. Once the $R$ matrix is obtained, the phase shifts for the spin-singlet and non-coupled spin-triplet states can be written as:
\begin{equation}
	\tan\delta^{0j}_{jj}(T_{lab})=-\frac{\pi}{2}qMR^{0j}_{jj}(q,q),
\end{equation}
\begin{equation}
	\tan\delta^{1j}_{jj}(T_{lab})=-\frac{\pi}{2}qMR^{1j}_{jj}(q,q).
\end{equation}
Since there is a tensor component in the $NN$ interaction, orbital angular momentum couples when $s=1$. For such coupled states, a unitary matrix exists that diagonalizes the $2\times2$ $R$ matrix. Thus, besides the phase shifts for the two coupled channels, there is a so-called mixing angle $\widetilde{\epsilon}_j$. Typically, in $NN$ scattering calculations, two labeling methods are commonly used. The first one, introduced by Blatt and Biedenharn \cite{blatt52}, involves eigenphase shifts $\widetilde{\delta}^j_{\mp}$ for the coupled channels, which can be expressed using the on-shell $R$ matrix as,
\begin{equation}\label{phs}
	\tan\widetilde{\delta}^j_{\mp}=-\frac{\pi}{4}qM\left[R_{--}^{j}+R_{++}^{j}\pm\frac{R_{--}^{j}-R_{++}^{j}}{\cos2\widetilde{\epsilon}_{j}}\right].
\end{equation}
The mixing angle coefficient $\widetilde{\epsilon}_j$ is given by:
\begin{equation}\label{phm}
	\tan2\widetilde{\epsilon}_{j}(T_{lab})=\frac{2R_{+-}^{j}}{R_{--}^{j}+R_{++}^{j}}.
\end{equation}
All the $R$ matrix elements in the above expressions are parameterized by $(q',q)$, where $q$ is the on-shell momentum. The abbreviations used here are:
\begin{eqnarray}
	&&R^j_{++}=R^{1j}_{j+1,j+1},~~~R^j_{--}=R^{1j}_{j-1,j-1},\nonumber\\
	&&R^j_{+-}=R^{1j}_{j+1,j-1},~~~R^j_{-+}=R^{1j}_{j-1,j+1}.
\end{eqnarray}

Since the distinction between protons and neutrons is crucial for high-precision nuclear forces, it's necessary to consider this distinction in $pp$, $nn$, and $np$ scattering. For $pp$ scattering, the relation between on-shell momentum $q^2$ and nucleon mass $M$ is:
\begin{equation}
	q^2=\frac{1}{2}M_{p}T_{lab},~~~M=M_{p},
\end{equation}
for $nn$ scattering:
\begin{equation}
	q^2=\frac{1}{2}M_{n}T_{lab},~~~M=M_{n},
\end{equation}
and for $np$ scattering:
\begin{equation}
	q^2=\frac{M_{p}^{2}T_{lab}(T_{lab}+2M_{n})}{(M_{p}+M_{n})^2+2T_{lab}M_{p}},~~~M=\frac{2M_pM_n}{M_p+M_n}.
\end{equation}
Here $M=M_{p}$ represents the proton mass, $M=M_{n}$ for the neutron mass, and $T_{lab}$ is the kinetic energy of the incident particle in the laboratory system, its relation with $q^2$ is based on relativistic kinematics. Another representation of $NN$ scattering phase shifts is proposed by Stapp et al~\cite{stapp57}. Here, $\epsilon_{j}$ and $\delta^{j}$ are used, which are related to the Blatt-Biedenharn phase shifts as:
\begin{eqnarray}
	\delta_{+}^{j}+\delta_{-}^{j}&=&\widetilde{\delta}_{+}^{j}+\widetilde{\delta}_{-}^{j},\nonumber\\
	\sin(\delta_{-}^{j}-\delta_{+}^{j})&=&\tan2\epsilon_{j}/\tan2\widetilde{\epsilon}_{j},\nonumber\\
	\sin(\widetilde{\delta}_{-}^{j}-\widetilde{\delta}_{+}^{j})&=&\sin2\epsilon_{j}/\sin2\widetilde{\epsilon}_{j}.
\end{eqnarray}
These expressions consider only $NN$ interaction and are suitable for $nn$ and $np$ scattering. For $pp$ scattering, the Coulomb interaction needs to be taken into account. 

The Lippmann-Schwinger equation is numerically solved using the matrix inversion method. A total of $6$ sets ($1$7 types) of high-precision realistic nuclear forces are calculated for various partial waves with laboratory incident energies less than $300$ MeV and total angular momentum $j\leq 3$. The $6$ sets of high-precision realistic nuclear forces are: the AV18 potential \cite{wiringa95}, which depends on spin-isospin operators in coordinate space; the CD-Bonn potential, based on the one-boson exchange model with charge dependence \cite{machleidt01}; the pvCD-Bonn potential, which considers pseudovector coupling between pion and nucleon based on the CD-Bonn potential \cite{wang19}; and three sets of chiral nuclear forces derived from chiral effective field theory, up to fifth-order momentum perturbation effects (N$^4$LO) \cite{epelbaum15a,entem17,reinert17}. The differences among these chiral nuclear forces mainly lie in the power counting rules for perturbation orders, the form of regularization factors, and the slightly different cutoff constants. {Of course, the uncertainties arising from different expansion orders in the chiral potentials are very important. Various methods to estimate these uncertainties are discussed in Refs.  \cite{epelbaum15a,entem17,furnstahl15,melendez17,melendez19} . In this work, we focus on comparing the high-precision 
$NN$ potentials from different theoretical frameworks. Thus, we will not include the uncertainties of chiral forces on phase shifts, cross sections, and so on in the results.}

All aforementioned high-precision realistic nuclear forces effectively reproduce experimental scattering data, achieving an accuracy of $\chi^2/N\sim 1$. Throughout our calculations, we meticulously consider the isospin effects of nucleon scattering, categorizing them into $pp$, $nn$, and $np$ channels. Among these, experimental pp scattering data currently boasts the highest accuracy, while $nn$ scattering data remains relatively sparse. We will sequentially present our computed results and analysis of phase shifts for $np$, $nn$, and $pp$ channels using the aforementioned high-precision nuclear forces.

\begin{figure*}[htb]
	\centering
	\includegraphics[width=1.0\textwidth]{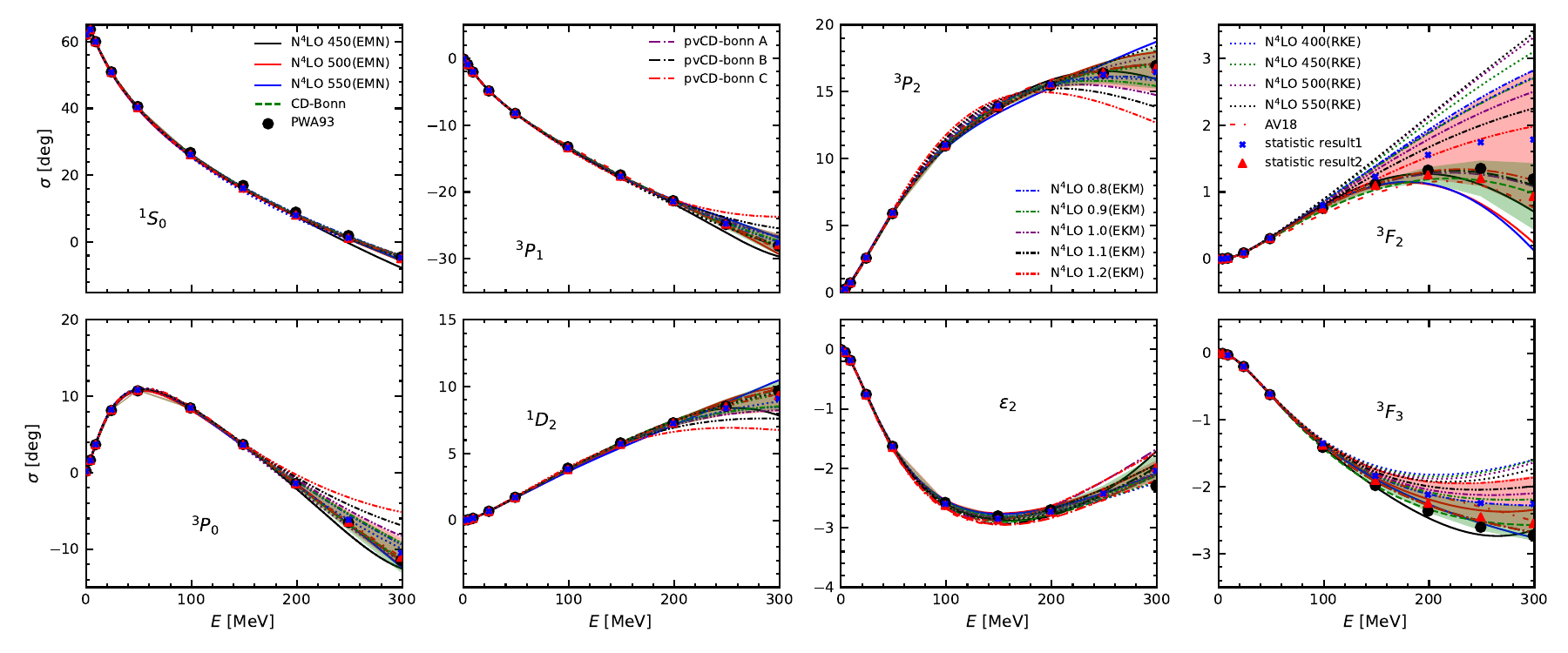}
	\caption{The variation of scattering phase shifts with incident energy for $np$ ($T=1$) channels calculated by various high-precision nuclear forces with $J\leq3$, along with statistical error analysis.}
	\label{fig:1}		
\end{figure*}

In Fig. \ref{fig:1}, we present the $np$ scattering phase shifts for the isospin $T=1$ channel calculated using 17 types of high-precision $NN$ interactions. These include AV18 \cite{wiringa95}, CD-Bonn \cite{machleidt01}, pvCD-Bonn A, B, C \cite{wang19}, N$^4$LO 450(EMN), N$^4$LO 500(EMN), N$^4$LO 550(EMN) \cite{entem17}, N$^4$LO 0.8(EKM), N$^4$LO 0.9(EKM), N$^4$LO 1.0(EKM), N$^4$LO 1.1(EKM), N$^4$LO 1.2(EKM) \cite{epelbaum15a}, N$^4$LO 400(RKE), N$^4$LO 450(RKE), N$^4$LO 500(RKE), and N$^4$LO 550(RKE) \cite{reinert17} potentials. Additionally, to compare with experimental data, we also include the phase shift values obtained by the Nijmegen group in 1993 through partial wave analysis (PWA93) \cite{stoks94} of scattering data.

For the $^1S_0$ channel, within the incident energy range up to $E\leq 300$ MeV, except for the chiral force N$^4$LO 450(EMN) from the Idaho group, the phase shifts obtained from the other $16$ high-precision nuclear forces are consistent, and they match completely with the PWA93 data. At $E\geq 200$ MeV, the deviation of the $^1S_0$ phase shift from N$^4$LO 450(EMN) compared to other potentials should mainly stem from the smaller cutoff value $\Lambda$. For channels with $l\leq2$, such as $^3P_0$, $^3P_1$, $^1D_2$, and $^3P_2$, all high-precision nuclear forces yield essentially identical phase shift values at $E\leq 200$ MeV. As the incident energy increases, deviations in the phase shifts computed using chiral forces begin to appear, which are also related to their smaller cutoff values. In $^3F_2$ and $^3F_3$ channels, these deviations are further amplified. Only when $E\leq 100$ MeV, the phase shift values from the $17$ potentials are relatively close. This is because in chiral effective theory, for N$^4$LO forces, the attempt to fit the experimental values for low-energy couplings is truncated only to the $^3D_3$ channel, so the phase shift values for $^3F_2$ and $^3F_3$ channels can only be regarded as theoretical predictions. Thus, various counting rules for different perturbation orders and cutoff momenta may affect the phase shift values for these two channels.

We assume that the phase shifts generated by these potentials at each incident energy point follow a Gaussian distribution, 
\begin{equation}
	X\sim \mathcal{N}(\mu,\sigma^2)=\frac{1}{\sigma\sqrt{2\pi}}e^{\left(\frac{x-\mu}{\sigma}\right)^2},
\end{equation}
where $\mu$ is the mean value and $\sigma$ is the standard deviation. Therefore, we can consider $\mu\pm\sigma$ as the central value and its deviation for theoretical phase shift calculations. The shaded areas in Fig. \ref{fig:1} represent these results. Due to significant discrepancies between the predictions of many chiral forces and the results of partial wave analysis in describing $F$-channel phase shifts, when performing statistical analysis on the theoretical phase shift results, we selected two groups of potentials: the first group includes all $17$ potentials (shaded in red), while the second group excludes the $9$ potentials from N$^4$LO (EKM) and N$^4$LO (RKE) (shaded in green).

We found that for channels with $l<3$, the choice of potential sets had minimal impact on the final central values and standard deviations. However, for the $F$-wave channels, excluding the results from the EKM and RKE sets of chiral nuclear forces significantly reduced the standard deviation of the second group, and the central values remained consistent with the results from PWA93. Therefore, caution should be exercised when using the results from these two sets of chiral nuclear forces, especially for energies greater than $200$ MeV. Particularly, for the $P$-wave and $D$-wave channels, significant discrepancies were observed at higher incident energies compared to the results obtained with other potential sets, especially when the truncation length of the EKM potential is relatively large.

\begin{table*}[htb]
	\centering
	\scalebox{0.8}{
		\begin{tabular}{ccccccccc}
			\hline				
			$E$/MeV& PWA93&N$^4$LO500 (EMN)&CD-Bonn&pvCD-Bonn B&N$^4$LO1.0 (EKM)&N$^4$LO400 (RKE)&AV18&Analysis\\
			\hline
			1&	62.07&	62.03&	62.11&	62.13&	62.10&	62.07&	61.80&	62.07±0.07\\
			5&	63.63&	63.47&	63.68&	63.69&	63.65&	63.58&	63.39&	63.61±0.08 \\
			10&	59.96&	59.72&	60.01&	60.02&	59.97&	59.86&	59.69&	59.92±0.10 \\
			25&	50.90&	50.47&	50.93&	50.93&	50.90&	50.66&	50.55&	50.80±0.14\\
			50&	40.54&	39.82&	40.44&	40.43&	40.43&	39.99&	40.03&	40.30±0.20\\
			100&26.78&	25.66&	26.37&	26.32&	26.34&	25.50&	25.97&	26.19±0.30\\
			150&16.94&	15.76&	16.30&	16.24&	16.22&	15.17&	15.94&	16.11±0.36\\
			200&8.94&	7.88&	8.29&	8.23&	8.15&	7.15&	7.95&	8.06±0.37\\
			250&1.96&	0.93&	1.56&	1.51&	1.41&	0.77&	1.24&	1.27±0.48\\
			300	&-4.46&	-5.60&	-4.28&	-4.31&	-4.35&	-4.31&	-4.58&	-4.62±0.84\\				
			\hline				
	\end{tabular}}
	\caption{The phase shifts of various interaction potentials in the $np$ $^{1}S_{0}$ channel, along with their averages and statistical error analysis.}\label{table 1}
\end{table*}

In Table \ref{table 1}, we present specific values of scattering phase shifts obtained by various typical $np$ interaction potentials in the $^{1}S_{0}$ channel at different incident energies. The values in the second column are derived from the partial wave analysis PWA93. The intermediate columns correspond to the results obtained from $\mathrm{N^{4}LO}$ 500(EMN), CD-Bonn, pvCD-Bonn B, $\mathrm{N^{4}LO}$ 1.0(EKM), $\mathrm{N^{4}LO}$ 400(RKE), and AV18 potentials, respectively. The last column represents the central values and deviations of the phase shifts obtained using Gaussian distribution for all $17$ potentials. The deviations are relatively small at lower incident energies but gradually increase with energy. Particularly, the scattering phase shift of $\mathrm{N^{4}LO}$ 500(EMN) at an incident energy of $300$ MeV is significantly smaller than that of other potentials. This results in a slightly smaller central value obtained through statistical analysis compared to the data from PWA93. However, considering the effect of deviations, this result is reasonable.

\begin{figure*}[htb]
	\centering
	\includegraphics[width=1.0\textwidth]{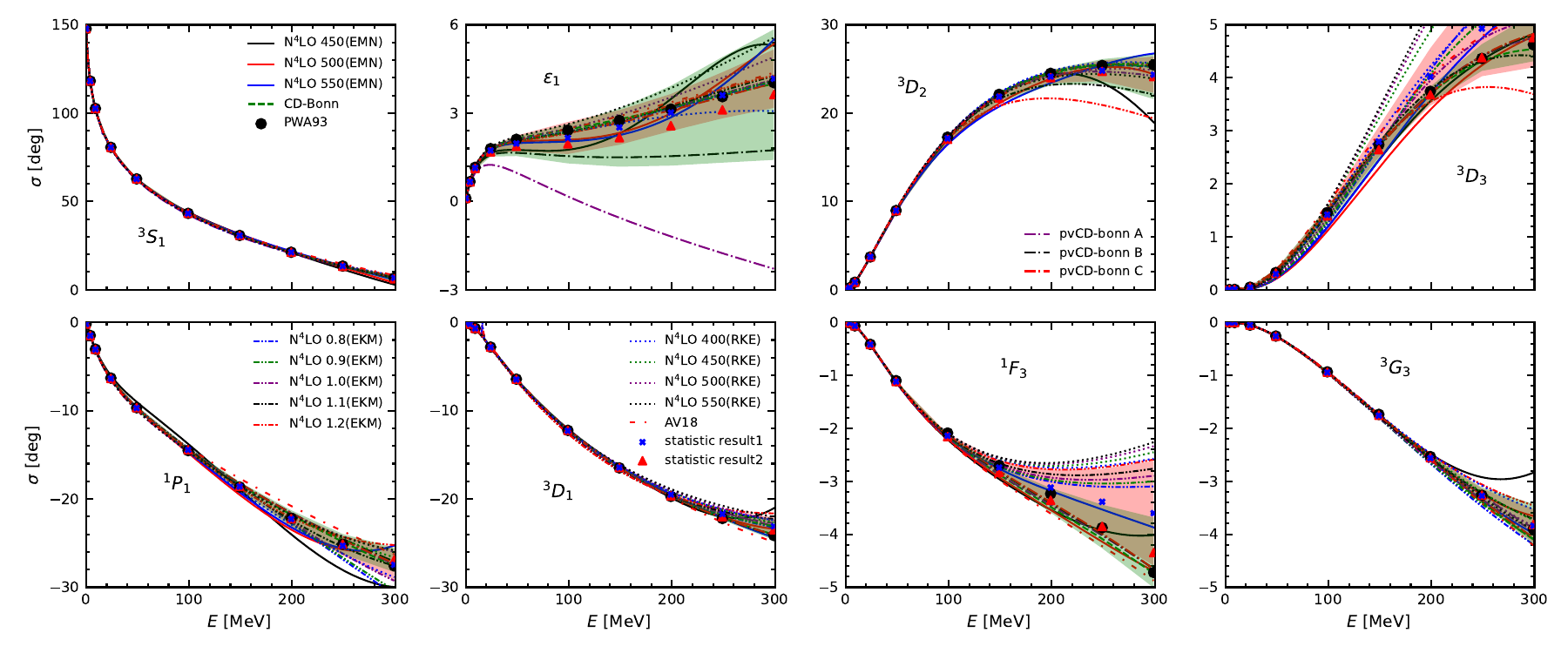}
	\caption{The $np$ ($T=0$) channel's phase shift as a function of incident energy, computed by various high-precision nuclear forces for $j\leq3$, along with a statistical error analysis, is depicted in the graph.}
	\label{fig:2}
\end{figure*}

In $np$ interactions, besides the isovector channel ($T=1$), there is also the case of the isoscalar channel ($T=0$). Nuclear potentials involve a strong tensor force component, primarily manifested in the mixing effect of the $^{3}S_{1}$ and $^{3}D_{1}$ channels in $np$ interactions. This is closely related to the mixing angle $\varepsilon_1$. The nuclear force in the $^{3}S_{1}$ channel exhibits attraction throughout the entire range of incident energies, with phase shift values obtained from various interactions being essentially consistent. However, for other channels, as the incident energy increases, differences in the phase shift values calculated by various high-precision nuclear forces become increasingly apparent, especially when the incident energy exceeds $100$ MeV. The differences in the mixing angles $\varepsilon_1$ calculated by these nuclear forces are also significant, indicating variations in their tensor components. For instance, the trend of the mixing angle $\varepsilon_1$ for the pvCD-Bonn A potential changes opposite to other potentials as the energy increases, showing an initial increase followed by a decrease. For $j=3$ channels, the computed values from various chiral nuclear forces significantly deviate from the PWA93 partial wave analysis. Therefore, when conducting statistical analysis on the phase shifts generated by these potentials, we adopt the same approach as for the isovector $T=1$ case. One approach considers all phase shifts generated by interactions, while the other excludes the $\mathrm{N^{4}LO}$(EKM) and $\mathrm{N^{4}LO}$(RKE) series of chiral  forces. The latter approach yields phase shift center values that are generally closer to the results of the PWA93 partial wave analysis.

\begin{table*}
	\centering
	\scalebox{0.8}{
		\begin{tabular}{ccccccccc}
			\hline
			$E$/MeV& PWA93&N$^4$LO500(EMN)&CD-Bonn&pvCD-Bonn B&N$^4$LO1.0(EKM)&N$^4$LO400(RKE)&AV18&Analysis\\
			\hline
			1&    147.75	&147.74	&147.74&	147.76&	147.75&	147.74&	148.00&	147.75±0.06\\
			5&	118.18&	118.17&	118.17&	118.20&	118.18&	118.15&	118.52&	118.18±0.09 \\
			10&	102.61&	102.60&	102.61&	102.63&	102.61&	102.56&	102.94	&102.60±0.10\\
			25&	80.63&	80.64&	80.62&	80.63&	80.60&	80.50&	80.93&	80.60±0.13\\
			50&	62.77&	62.89&	62.71&	62.70&	62.71&	62.49&	63.08&	62.71±0.19\\
			100&43.23&	43.70&	43.05&	43.02&	43.13&	42.68&	43.65&	43.16±0.36\\
			150&30.72&	31.40&	30.44&	30.43&	30.64&	30.11&	31.31&	30.67±0.42\\
			200	&21.22&	21.58&	20.93&	20.93&	21.20&	20.87&	22.04	&21.18±0.33\\
			250&13.39&	12.65&	13.19&	13.22&	13.52&	13.70&	14.54&	13.36±0.57\\
			300&6.60&	3.98&	6.62&	6.68&	7.04&	8.05&	8.21&	6.66±1.30\\				
			\hline								
	\end{tabular}}
	\caption{The $np$ $^{3}S_{1}$ channel: phase shifts of various interaction potentials and their average values with statistical error analysis.}\label{table 3}
\end{table*}

Table \ref{table 3} presents the numerical values of phase shifts for various interaction potentials in the $np$ $^{\mathrm{3}}S_{\mathrm{1}}$ channel, ranging from $1$ MeV to $300$ MeV of incident nucleon energy. The last column displays the phase shift values calculated considering all potential fields, along with the average phase shift obtained through statistical analysis and the corresponding statistical error at different energies. The positive phase shift values in the $^{\mathrm{3}}S_{\mathrm{1}}$ channel within this energy range indicate strong attractive interactions between nucleons. At energies below $200$ MeV, the phase shift values computed by various high-precision nuclear forces are generally consistent with the results of the PWA93. However, as the energy increases, deviations gradually appear in the results for different interactions, with the $\mathrm{N^{4}LO}$ 500(EMN) potential showing the largest discrepancy. From Table \ref{table 3}, it can be observed that the central value at 5 MeV incident nucleon energy is $147.75$  with a deviation of $0.06$, at $100$ MeV it is $43.16$ with a deviation of $0.36$, and at $300$ MeV it is $6.66$ with a deviation of $1.30$. 

Since neutrons are charge neutral, there is currently very little experimental data on $nn$ scattering, so there are no results from phase shift analyses. However, the charge-independent symmetry of nuclear forces has a weak breaking, so there is only a slight difference between the scattering phase shifts of $nn$ and $np$ channels in the $T=1$ channel. Especially at lower incident energies, due to the Pauli exclusion principle, the attractive strength of $nn$ interaction is slightly smaller compared to the $np$ interaction in the $^{\mathrm{1}}S_{\mathrm{0}}$ channel.

\begin{figure*}
	\centering
	\includegraphics[width=1.0\textwidth]{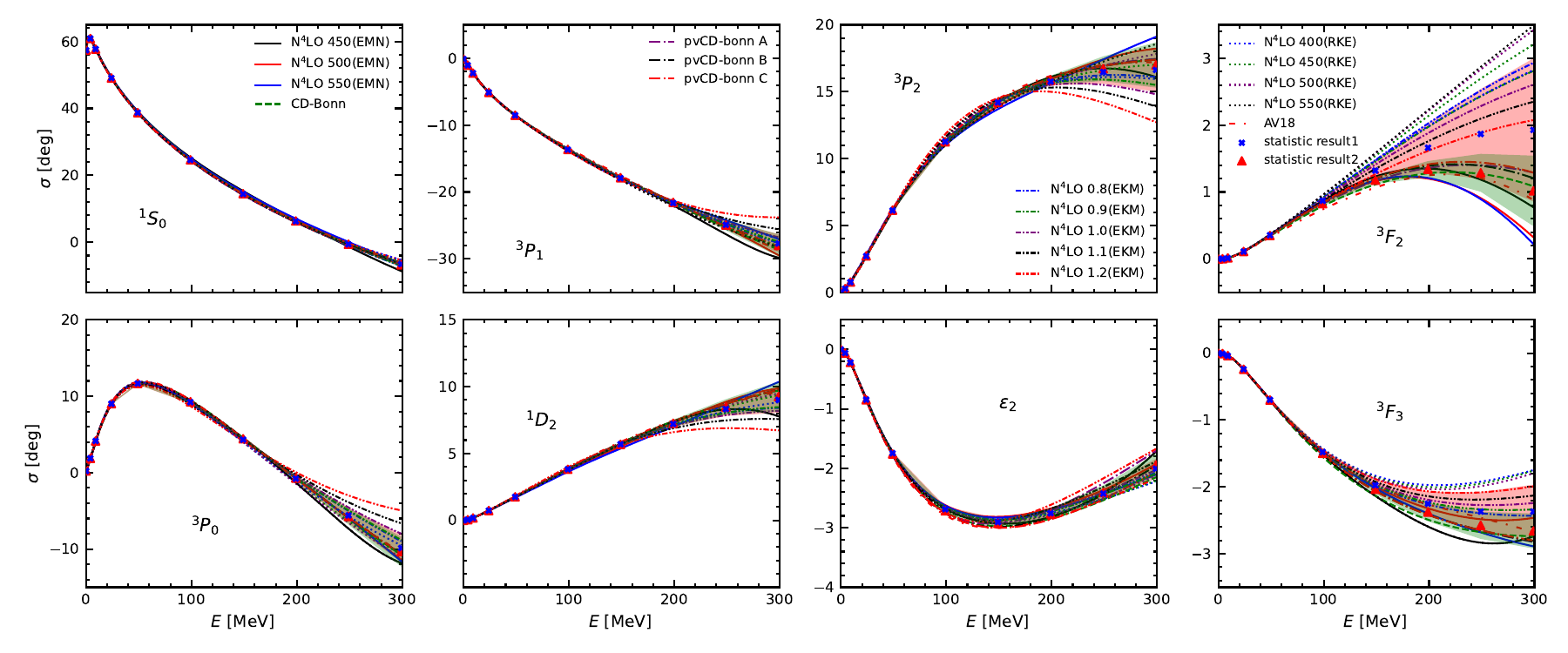}
	\caption{The variation of scattering phase shifts with incident energy for $nn$ ($T=1$) channels calculated by various high-precision nuclear forces with $J\leq3$, along with statistical error analysis.}
	\label{fig:3}		
\end{figure*}

In Fig.~\ref{fig:3}, the phase shifts of $nn$ scattering from $17$ types of high-precision nuclear potential are shown with the  corresponding statistical mean values and deviations. They have quite similar tendencies to those of $np$ scattering. Table \ref{table 5} lists the $^{\mathrm{1}}S_{\mathrm{0}}$ channel shifts calculated using $\mathrm{N^{4}LO}$ 500(EMN), CD-Bonn, pvCD-Bonn B, $\mathrm{N^{4}LO}$ 1.0(EKM), $\mathrm{N^{4}LO}$ 400(RKE), and AV18 potentials. It can be observed that the range of shifts in the $^{\mathrm{1}}S_{\mathrm{0}}$ channel is from $60$ to $-7$. The shifts from $1$ MeV to $50$ MeV within the range of nucleon incident energy  are consistent across various potentials. However, deviations between different potentials calculated after $100$ MeV gradually increase, with $\mathrm{N^{4}LO}$ 1.0(EKM) having the largest deviation, while the others have relatively small deviations. 

\begin{table*}
	\centering
	\scalebox{0.8}{
		\begin{tabular}{cccccccc}
			
			\hline
			$E$/MeV&N$^4$LO500(EMN)&CD-Bonn&pvCD-Bonn B&N$^4$LO1.0(EKM)&N$^4$LO400(RKE)&AV18&Analysis\\
			\hline
			1&	57.48&	57.48&	57.25&	57.36&	57.45&	57.49&	57.40±0.09\\
			5&	60.94&	60.92&	60.83&	60.88&	60.98&	60.88&	60.91±0.06\\
			10&	57.78&	57.74&	57.69&	57.73&	57.84&	57.67&	57.76±0.07\\
			25&	49.08&	49.02&	49.03&	49.07&	49.19&	48.94&	49.10±0.10\\
			50&	38.70&	38.60&	38.64&	38.70&	38.78&	38.59&	38.73±0.14\\
			100&	24.65&	24.38&	24.44&	24.50	&24.45&	24.55&	24.58±0.24\\
			150&14.71&	14.14&	14.21&	14.24&	14.15&	14.48&	14.39±0.32\\
			200	&6.76&	5.97&	6.06&	6.05&	6.15&	6.44&	6.24±0.30\\
			250	&-0.26&	-0.91&	-0.79&	-0.79&	-0.22&	-0.33&	-0.63±0.35\\
			300&	-6.85&	-6.88&	-6.74&	-6.60&	-5.29&	-6.21&	-6.57±0.73\\				
			\hline				
	\end{tabular}}
	\caption{The phase shifts of various interaction potentials in the $nn$ $^{1}S_{0}$ channel, along with their averages and statistical error analysis.}\label{table 5}
\end{table*}

\begin{figure*}
	\centering
	\includegraphics[width=1.0\textwidth]{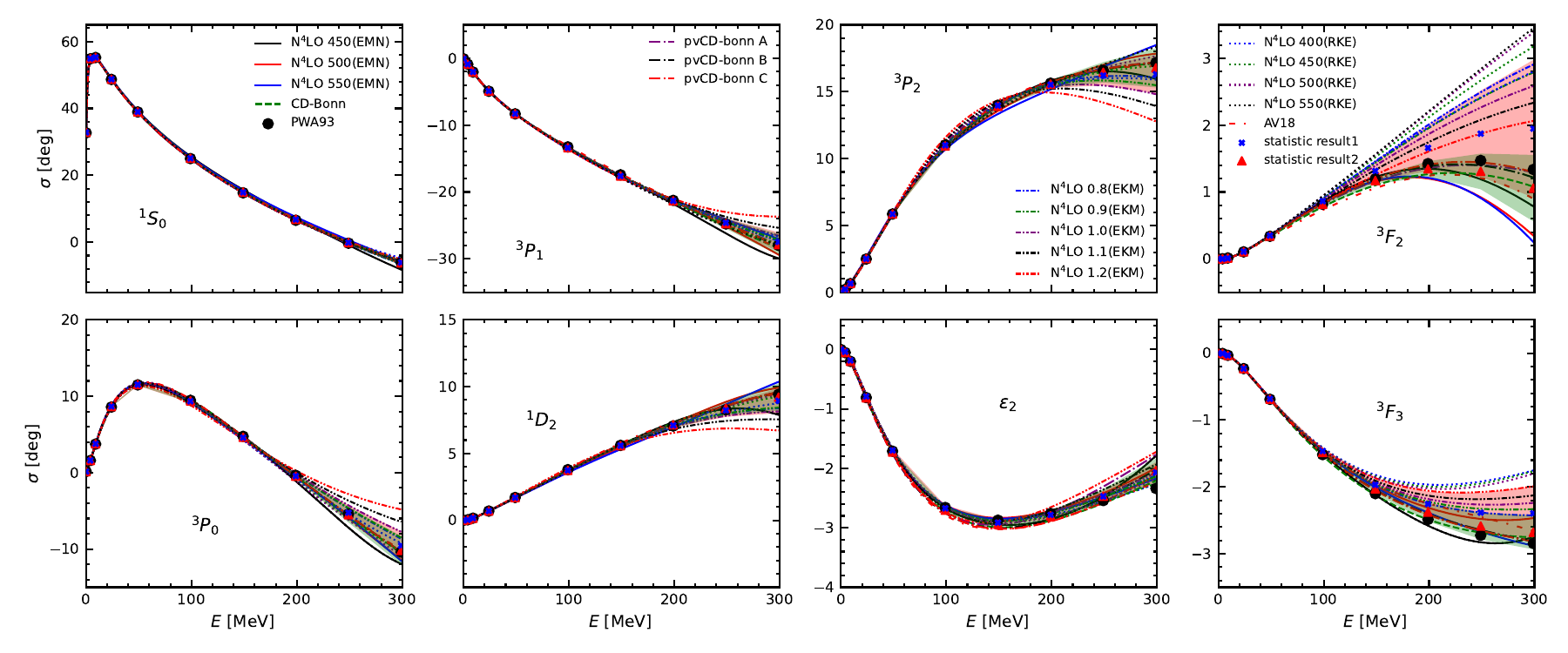}
	\caption{The variation of scattering phase shifts with incident energy for pp ($T=1$) channels calculated by various high-precision nuclear forces with $J\leq2$, along with statistical error analysis.}
	\label{fig:4}	
\end{figure*}

Fig. \ref{fig:4} presents the pp channel scattering cases with the same potentials as the $nn$ and $np$ cases. For $pp$ potential, a Coulomb repulsive interaction should be considered. Therefore, the phase shifts of $pp$ scattering are reduced at low incident energy regions compared to those of $nn$ and $np$ scattering. The Coulomb force is a long-range interaction, which has small influence on the high-energy scattering region. These high-precision nucleon potentials exhibit similar behaviors as in the $nn$ and $np$ cases. 

Table \ref{table 7} presents the phase shifts in the $pp$ $^{\mathrm{1}}S_{\mathrm{0}}$ channel. It can be observed from the table that the range of shifts in the $^{\mathrm{1}}S_{\mathrm{0}}$ channel is from $56$ to $-7$ in the incident energy range from $1$ MeV to $100$ MeV. However, deviations gradually increase as the energy rises, with $\mathrm{N^{4}LO}$ 500(EMN) having the largest deviation, while others are closer to the mean values. According to the central value and deviation calculations, at $5$ MeV, the central value is $54.91$ with a deviation of $0.05$, at $100$ MeV, the central value is $25.06$ with a deviation of $0.2$, and at $300$ MeV, the central value is $-6.07$ with a deviation of $0.73$. The uncertainties from these potentials in the $pp$ channel are smallest.
\begin{table*}
	\centering
	\scalebox{0.8}{
		\begin{tabular}{ccccccccc}
			
			\hline
			$E$/MeV& PWA93&N$^4$LO500(EMN)&CD-Bonn&pvCD-Bonn B&N$^4$LO1.0(EKM)&N$^4$LO400(RKE)&AV18&Analysis\\
			\hline
			1&	32.68&	32.88&	32.88&	32.88&	32.79&	32.79&	32.88&	32.83±0.06\\
			5&	54.83&	54.93&	54.93&	54.95&	54.86&	54.92&	54.88&	54.91±0.05\\
			10&	55.22&	55.26&	55.27&	55.30&	55.22&	55.32&	55.17&	55.27±0.06\\
			25&	48.67&	48.66&	48.67&	48.72&	48.68&	48.82&	48.56&	48.72±0.09\\
			50&	38.93&	38.86&	38.89&	38.94&	38.94&	39.07&	38.84&	38.98±0.12\\
			100&	24.99&	24.99&	24.92&	24.97&	24.98&	25.00&	25.06&	25.06±0.20\\
			150&	14.76&	15.06&	14.74&	14.78&	14.75&	14.74&	15.04&	14.90±0.28\\
			200	&6.55&	7.11&	6.59&	6.63&	6.56&	6.73&	7.03&	6.76±0.28\\
			250&	-0.31&	0.11&	-0.28&	-0.23&	-0.29&	0.33&	0.27	&-0.12±0.35\\
			300&	-6.15&	-6.44&	-6.25&	-6.18&	-6.15&	-4.79&	-5.61&	-6.07±0.73\\			
			\hline			
	\end{tabular}}
	\caption{The phase shifts of various interaction potentials in the $pp$ $^{1}S_{0}$ channel, along with their averages and statistical error analysis.}\label{table 7}
\end{table*}

\section{The cross section of nucleon-nucleon scattering} \label{sec3}
Once the phase shifts are obtained, the $S$ matrix related to the scattering amplitude can be constructed as for the uncoupled channels $l=l'=j$ \cite{arriola20},
\begin{eqnarray}
	S^{sj}_{ll'}=e^{2i\delta^{sj}_{jj}},
\end{eqnarray}
and for the coupled channels $l=j\pm1,~l'=j\pm1$,
\begin{equation}
	S^{sj}_{ll'}=\begin{pmatrix}
		e^{2i\delta^{j}_{-}}\cos 2\epsilon_j & 	ie^{i(\delta^{j}_{-}+\delta^{j}_{+})}\sin 2\epsilon_j	\\
		ie^{i(\delta^{j}_{-}+\delta^{j}_{+})}\sin 2\epsilon_j &	e^{2i\delta^{j}_{+}}\cos 2\epsilon_j
	\end{pmatrix} .
\end{equation}
The corresponding scattering amplitude for the $NN$ system at the scattering angle $\theta$ is 
\begin{eqnarray}\label{smt}
	M^s_{m',m}(\theta)&=&\frac{1}{2ik}\sum_{j,l',l}i^{l-l'}\sqrt{4\pi(2l+1)}Y_{l',m'-m}(\theta,0)C(l'js|m-m',m',m)\\\nonumber
	&\times&C(ljs|0,m,m)(S^{sj}_{ll'}-\delta_{l',l}),
\end{eqnarray}
where $C$ is the Clebsch-Gordan coefficients. This scattering amplitude is given in the spin-coupled representation, whereas it is more convenient to show it in the helicity basis for describing the $NN$ scattering observables,
\begin{equation}
	M_{m'_1,m'_2,m_1,m_2}=\sum_{m,m',s,s'}C\left(\frac{1}{2}\frac{1}{2}s|m_1,m_2,m\right)C\left(\frac{1}{2}\frac{1}{2}s|m'_1,m'_2,m'\right)M^s_{m,m'}.
\end{equation} 
$m_1,~m_2,~m'_1,~m'_2=\pm\frac{1}{2}$ are the third spin components of input and scattering nucleons, respectively. Therefore, the scattering amplitude is a $4\times4$ matrix for the $NN$ system. Due to the parity and time-reversal symmetries, there are only $5$ independent complex components in this matrix, which can be parameterized. Many attempts and notations were developed for this parameterization process \cite{wolfenstein56,hoshizaki69,bystricky79}. A traditional definition is the Wolfenstein scheme \cite{wolfenstein56},
\begin{eqnarray}
	M(\hat{\vec k}_f,\hat{\vec k}_i)&=&a+m(\sigma_1\cdot \vec n)(\sigma_2\cdot \vec n)+(g-h)(\sigma_1\cdot \vec m)(\sigma_2\cdot \vec m)\\\nonumber
	&+&(g+h)(\sigma_1\cdot \vec l)(\sigma_2\cdot  l)+c(\sigma_1+\sigma_2)\cdot \vec n
\end{eqnarray}
where, $\vec n,~\vec m,~\vec l$ are the three unitary vectors along the directions of $\vec k_f\perp\vec k_i$, $\vec k_f-\vec k_i$, and $\vec k_f+\vec k_i$.

For the spin-unpolarized input and scattering states, the differential cross section of $NN$ scattering can be presented by these five Wolfenstein parameters,
\begin{eqnarray}\label{dsg}
	\frac{d\sigma}{d\Omega}&=&\frac{1}{4}\sum_{m'_1,m'_2,m_1,m_2}|M_{m'_1,m'_2,m_1,m_2}(\hat{\vec k}_f,\hat{\vec k}_i)|^2=\frac{1}{4}\text{Tr}\left[{MM^\dagger}\right]\\\nonumber
	&=&|a|^2+|m|^2+2|c|^2+2|g|^2+2|h|^2.
\end{eqnarray}
The spin-polarized observables can also be derived with the spin density matrix operator and its polarization projection operator. 

Once the phase shifts at different partial wave channels are obtained, the scattering observables can be generated by the scattering amplitude given by Eq. (\ref{smt}). It was parameterized as $a,~c,~m,~g,~h$ by the Wolfenstein scheme. In the present work, the differential cross sections $d\sigma/d\Omega$ in Eq. (\ref{dsg}) from different potentials are evaluated, where the angular partial waves must be included up to $j=20$. 

\begin{figure*}
	\centering
	\includegraphics[width=0.5\textwidth]{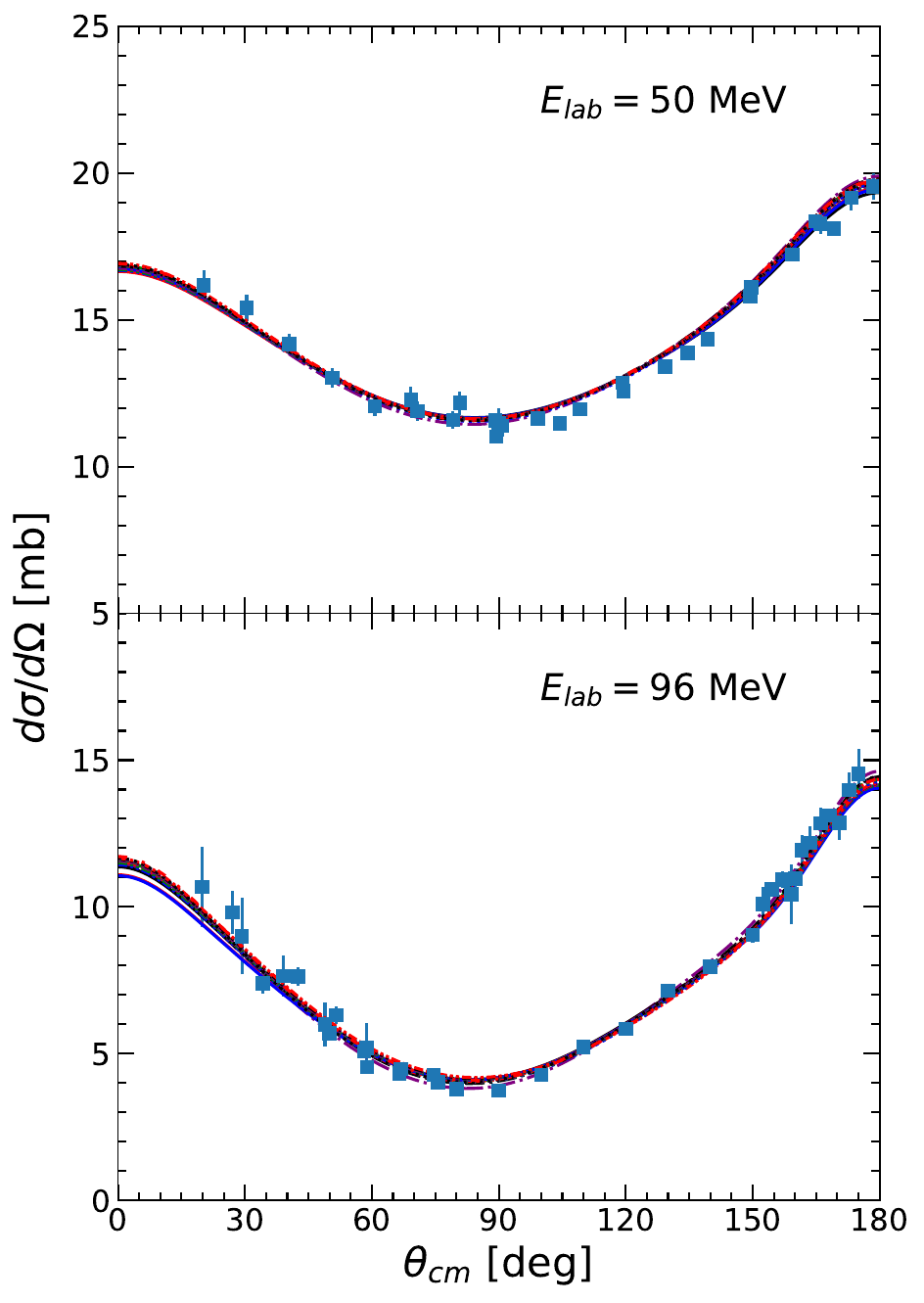}
	\caption{The differential cross sections for $pp$ scattering calculated by various high-precision nuclear forces with incident energies, $E_{lab}=50$ MeV and $E_{lab}=96$ MeV. The meanings of line styles are same as Fig. \ref{fig:4}. }
	\label{fig:5}	
\end{figure*}

In Fig.~\ref{fig:5}, the differential cross section as a function of center-of-mass scattering angle with laboratory energies at $E_{lab}=50$ MeV and $E_{lab}=96$ MeV are plotted. The corresponding experimental data are compared.  The 17 types of high-precision nuclear potentials provide the identical differential cross section at $E_{lab}=50$ MeV, which is consistent with the experimental data very well.  When the $E_{lab}$ grows up to $96$ MeV, the results from  $\mathrm{N^{4}LO}$ 500(EMN) and pvCD-Bonn A have significant differences from other calculations. The $d\sigma/d\Omega$ at small $\theta_{cm}$ of $\mathrm{N^{4}LO}$ 500(EMN) becomes lower. The one from pvCD-Bonn A reduces round $\theta_{cm}=90$ and is closer to the experimental data, whose mixing angle has a distinct behavior compared to other potentials.

\begin{figure*}
	\centering
	\includegraphics[width=0.5\textwidth]{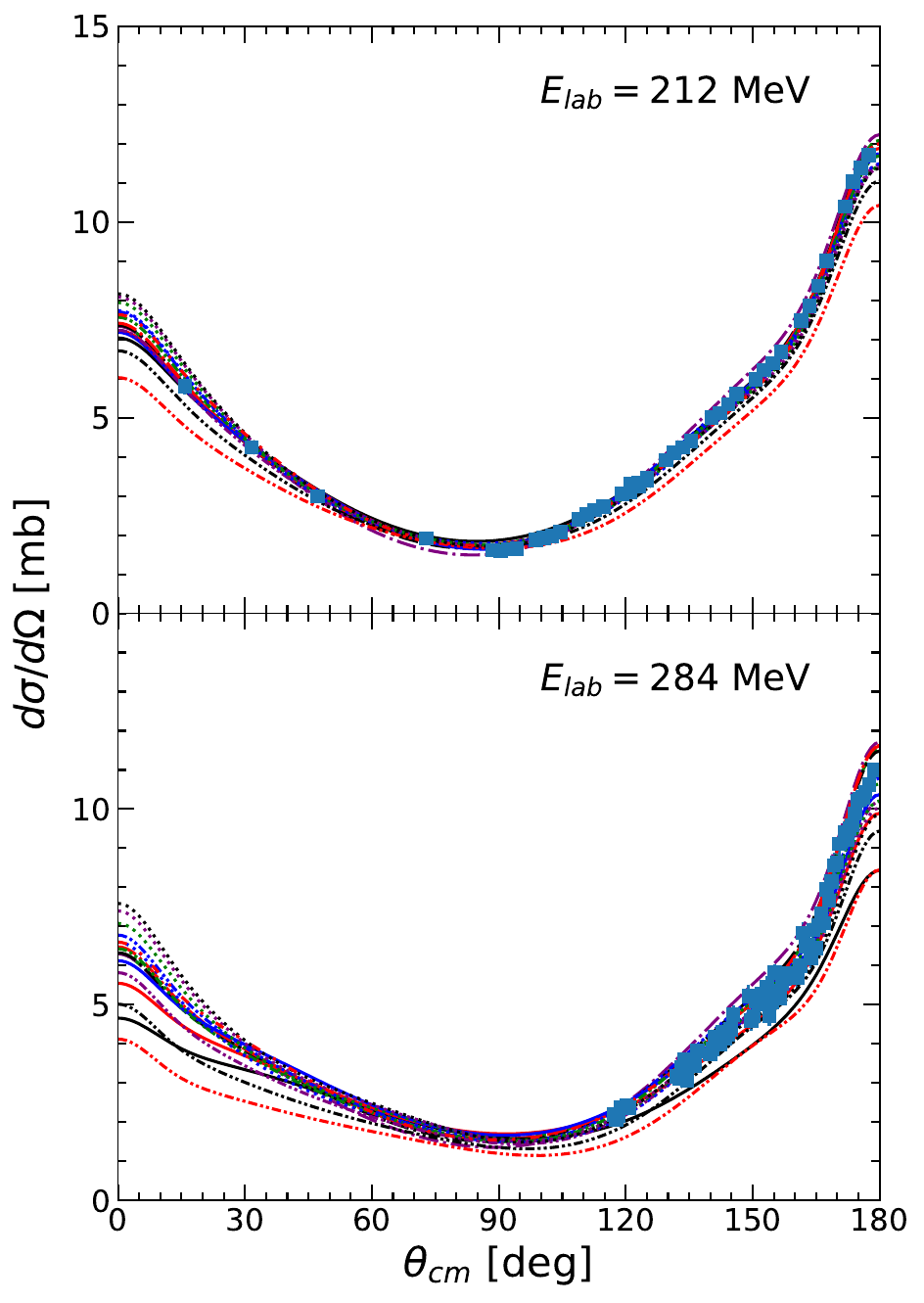}
	\caption{The differential cross sections for $pp$ scattering calculated by various high-precision nuclear forces with incident energies, $E_{lab}=212$ MeV and $E_{lab}=284$ MeV. The meanings of line styles are same as Fig. \ref{fig:4}.}
	\label{fig:6}
\end{figure*}

With the incident energy increment, the discrepancies in $d\sigma/d\Omega$ from these potentials more pronounced. In Fig.~\ref{fig:6}, the differential cross sections for $pp$ scattering with $E_{lab}=212$ MeV and $E_{lab}=284$ MeV at different scattering angles are given. The chiral potentials, which do not accurately describe the $F$ channels, exhibit significant deviations from experimental data at both forward and backward scattering angles. This is due to the construction of the $S$-matrix based on phase shifts. At higher collision energies, the convergence of the momentum expansion for chiral potentials deteriorates. 

\section{Entanglements of the nucleon-nucleon scattering}	\label{sec4}
The entanglement of identical particles perfectly exhibits the nonlocality of quantum mechanics and has attracted numerous interesting investigations in the field of quantum information in the past decades. Recently, it was applied to study the Wigner SU(4) symmetry in the strong interaction from the perspective of $NN$ scattering in the framework of chiral effective field theory. An entanglement power of $S$ matrix was defined to describe the averaged entanglement ability of the scattering matrix by considering all initial states \cite{beane19},
\begin{equation}
	\varepsilon (\hat S)=1-\int\frac{d\Omega_1}{4\pi}\frac{d\Omega_2}{4\pi}\text{Tr}_1[\hat{
		\rho}^2_1].
\end{equation}
Here, $\hat{\rho}_1$ is a reduced density matrix for the nucleon $1$ in the spin space which can be obtained from the two-nucleon density matrix $\hat{\rho}_{12}$,
\begin{equation}
	\hat{\rho}_1=\text{Tr}_2(\hat{\rho}_{12})=\text{Tr}_2(|\psi_\text{out}\rangle\langle\psi_\text{out}|).
\end{equation}
The scattering state $|\psi_\text{out}\rangle$ is generated by the initial state and $S$ matrix,
\begin{equation}
	|\psi_\text{out}\rangle=\hat{S}|\psi_\text{in}\rangle.
\end{equation}
The initial state can be represented by two angles in the spherical coordinate,
\begin{equation}
	|\psi_\text{in}\rangle=\left(\cos\frac{\theta_1}{2}, \exp({i\phi_1})\sin\frac{\theta_1}{2}\right)^T\otimes\left(\cos\frac{\theta_2}{2}, \exp({i\phi_2})\sin\frac{\theta_2}{2}\right)^T,
\end{equation}
where $d\Omega_1=\sin\theta_1d\theta_1d\phi_1$, $d\Omega_2=\sin\theta_2d\theta_2d\phi_2$ and $\theta_1, ~\theta_2\in[0,\pi],~\phi_1, ~\phi_2\in[0,2\pi]$. In the $l=0$ channel, the $S$-matrix is decomposed into the $^1S_0$ and $^3S_1$ channel, if the mixing angle between $^3S_1$ and $^3D_1$ is neglected. The $S$-matrix can be denoted as the phase shifts at singlet and triplet channels, $\delta_0$ and $\delta_1$,
\begin{equation}
	\hat S=\frac{1}{4}(e^{2i\delta_0}+3e^{2i\delta_1})\bm{1}-\frac{1}{4}(e^{2i\delta_0}-e^{2i\delta_1})\vec\sigma_1\otimes\vec\sigma_2.
\end{equation}
Therefore, the density matrix $\hat{\rho}_{12}$ in the spin space is $4\times 4$. Its reduced density matrix for the nucleon $1$ is reduced as $2\times 2$. The function $\text{Tr}_1[\hat{\rho}^2_1]$ is just related to $\theta_1,~\theta_2,~\phi_1$ and $\phi_2$. The entanglement power can be analytically written as,
\begin{equation}
	\varepsilon (\hat S)=\frac{1}{6}\sin^2[2(\delta_1-\delta_0)].
\end{equation}

\begin{figure*}
	\centering
	\includegraphics[width=0.9\textwidth]{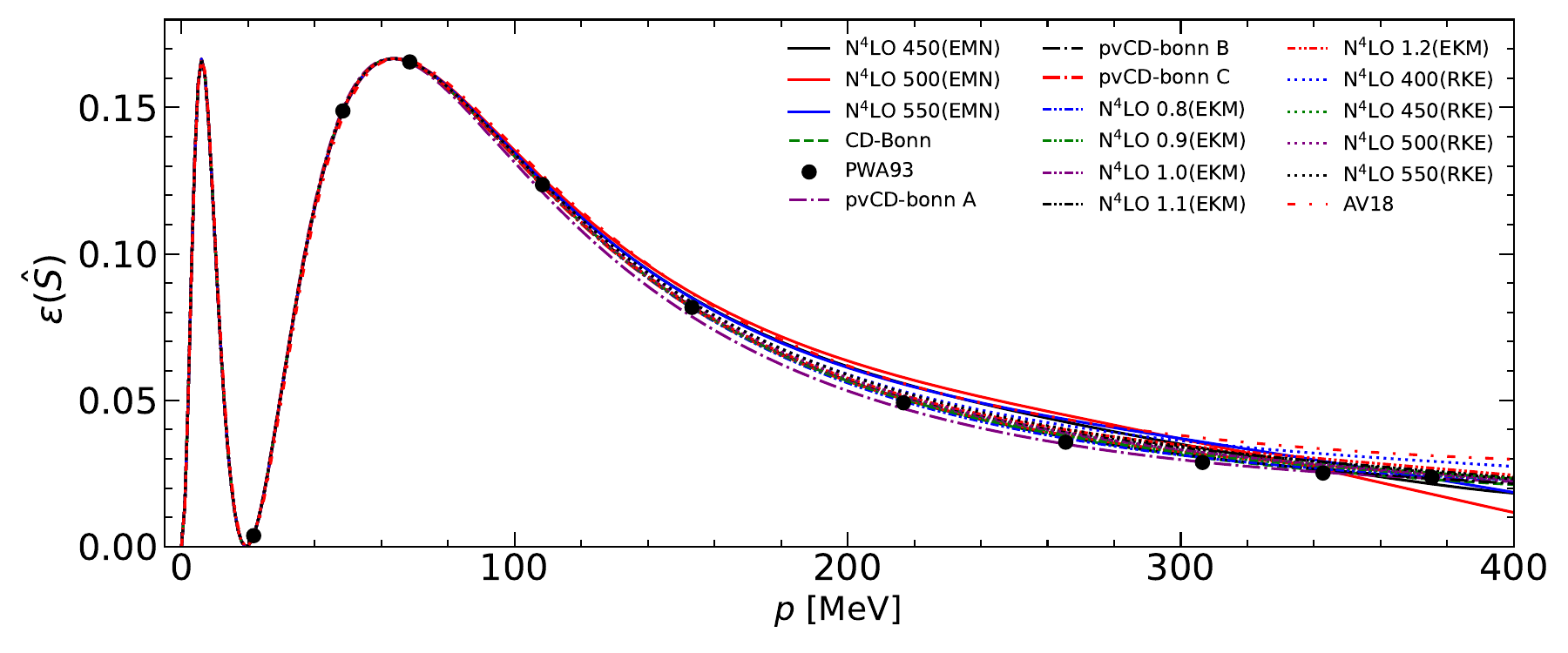}
	\caption{The entanglement power at $l=0$ channels from different high precision $NN$ potentials. These results are compared to those from PWA93.}
	\label{fig:7}
\end{figure*}

The wave function of $np$ scattering at tripled state $S=1$ can be considered as an entanglement state in the spin space. An entanglement power about $S$-matrix can be evaluated through the two-body scattering wave function, which can be represented by the phase shifts at singlet and triplet channels, $\delta_0$ and $\delta_1$ if the mixing angle $\varepsilon_1$ is neglected. In the Fig. \ref{fig:7}, the entanglement powers $\varepsilon(\hat{\mathcal{E}})$ from $17$ high precision $NN$ potentials are plotted and compared to that from PWA93. All of them equal $zero$ around input momentum $p=19.4$ MeV, where the phase shifts have a difference with $\pi/2$ between singlet and triplet channels and $|n\uparrow,p\downarrow\rangle$ converts into $|n\downarrow,p\uparrow\rangle$. At $p$ around $90$ MeV, the magnitude of entanglement power reaches the maximum value and then slowly reduces. The differences of $\varepsilon(\hat{\mathcal{E}})$  among $17$ potentials start to become obvious after their peak values. In the intermediate momentum region, the entanglement from pvCD-Bonn A potential has the minimum value and is most consistent with that from the partial wave analysis of experimental data. When $p$ approaches $400$ MeV, the  $\varepsilon(\hat{\mathcal{E}})$ from $\mathrm{N^{4}LO}$ 500(EMN) shows a distinct difference compared to other potentials, since from Table \ref{table 3}, its phase shifts at $np$ $^3S_1$ channel are smaller than other one at high input energy region.  {Overall, the entanglement powers from the current high-precision $NN$ potentials show very small differences, as the phase shifts for the $^1S_0$ and $^3S_1$ states of $NN$ system are well-determined. In contrast, in other two-baryon systems, such as $\Sigma^+ p$ or $\Lambda p$, the entanglement powers from different baryon-baryon potentials exhibit distinct behaviors \cite{liu24}. Furthermore, the spin entanglement can be detected by measuring the spin polarizations of one nucleon in three directions during polarization experiments, which relates to the spin correlation parameters \cite{bai24}.}

\section{The Brueckner Hartree-Fock method} 	\label{sec5}
The realistic $NN$ interaction cannot be directly applied to exactly describe nuclear many-body systems, such as finite nuclei and infinite nuclear matter within the independent-particle approximation. It should be normalized to treat the strong repulsion at short-range and take three-body force into account in a non-relativistic framework.  An available renormalization method was named as  Brueckner Hartree-Fock (BHF) method, where an effective interaction $G$ will replace of the realistic $NN$ interaction, $V$ within the Bethe-Goldstone equation~\cite{hu17},
\begin{equation}
	G(\omega,k_F)=V+\sum_{k_1,k_2>k_F}V\frac{Q(k_1,k_2)}{\omega-\varepsilon_1(k_1)-\varepsilon_2(k_2)+i\epsilon}G(\omega,k_F),
\end{equation}
where $k_F$ is the Fermi momentum of the nucleon, $\omega$ is the starting energy, $Q(k_1,k_2)$ is the Pauli exclusion operator, and $\varepsilon(k)$ is the single-particle energy given as,
\begin{eqnarray}
	\varepsilon(k)&=&\frac{k^2}{2M_N}+U(k;k_F)\\\nonumber
	&=&\frac{k^2}{2M_N}+Re\sum_{k'\leq k_F}\langle kk'|G(\varepsilon_1(k_1)+\varepsilon_2(k_2), k_F)|kk'\rangle_A.
\end{eqnarray}
The subscript of $A$ indicates the anti-symmetry wave function. When the normalized interaction $G$ is applied in the Hartree-Fock approximation, the energy per nucleon of nuclear matter can be shown as,
\begin{equation}
	B/A=\frac{3k^2_F}{10M_N}+\frac{1}{2\rho}\sum_{k,k'\leq k_F}Re\langle kk'|G(\varepsilon_1(k_1)+\varepsilon_2(k_2), k_F)|kk'\rangle_A.
\end{equation} 

\begin{figure*}
	\centering
	\includegraphics[width=0.5\textwidth]{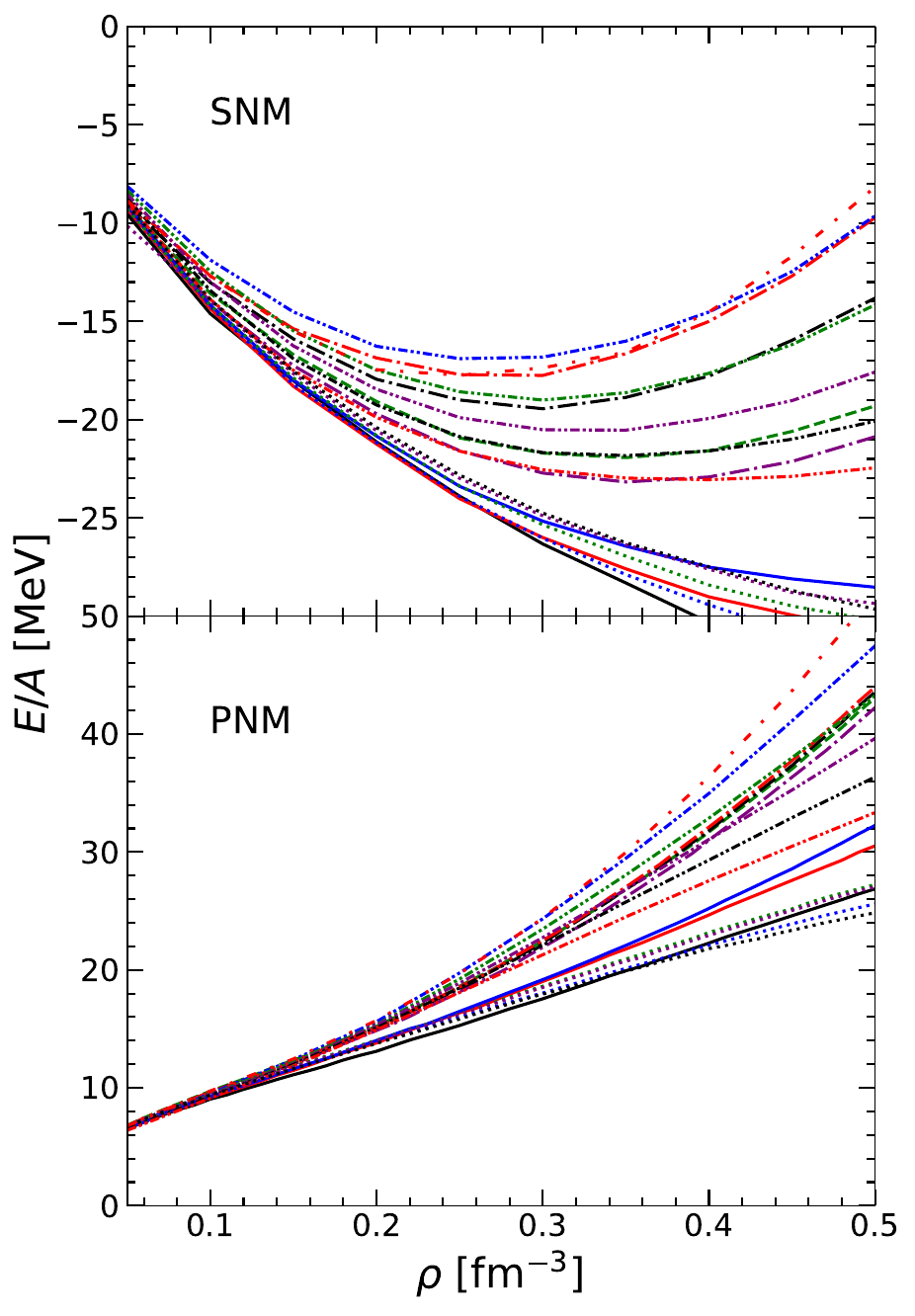}
	\caption{The energy per nucleon as a function of the density of symmetric nuclear matter and pure neutron matter with different $NN$ potentials in the framework of Brueckner-Hartree-Fock method. The meanings of line styles are same as Fig. \ref{fig:7}.}
	\label{fig:8}
\end{figure*}

In the last part, the EOS of nuclear matter is calculated in the framework of the BHF method, with different two-body high-precision $NN$ potentials. The symmetric nuclear matter case is shown in the upper panel of Fig. \ref{fig:8}. The binding energy per nucleon monotonically decreases with baryon density rising below $\rho<0.2$ fm$^{-3}$. The EOSs from $\mathrm{N^{4}LO}$ (EMN) and $\mathrm{N^{4}LO}$ (RKE) do not have the saturation character before  $\rho=0.5$ fm$^{-3}$, since they are very soft at short-range region. While another chiral potential, $\mathrm{N^{4}LO}$ (EKM) is regularized in the coordinate space, it becomes harder and can generate the saturation point. The other non-chiral potentials also can provide the saturation points around $\rho_{sat}=0.25$-$0.35$ fm$^{-3}$, which are a little far from the empirical data around $\rho_{sat}=0.16$ fm$^{-3}$. The repulsive contribution of three-body force is required in the non-relativistic framework to provide the reasonable saturation properties of symmetric nuclear matter.

The EOSs of pure neutron matter are plotted in the lower panel of  Fig. \ref{fig:8}. All of them monotonically increase with density and are almost identical below the empirical saturation density.  The differences become significant above $\rho=0.20$ fm$^{-3}$. The softer chiral potentials, $\mathrm{N^{4}LO}$ (EMN) and $\mathrm{N^{4}LO}$ (RKE), still have the lowest binding energy. The AV18 and $\mathrm{N^{4}LO}$ 0.8(EKM) contribute to the stiffer EOSs. Furthermore, it can be found that the regularized cut-off momentum in $\mathrm{N^{4}LO}$ (REK) has the smallest influence on the EOS, while that in $\mathrm{N^{4}LO}$ 0.8(EKM) has a very strong effect.

\section{Conclusions} \label{sec6}

In this work, the high-precision $NN$ interaction potentials proposed in recent years, including AV18 potential, CD-Bonn potential, chiral potentials, and pvCD-Bonn potentials,  were compared with each other from the aspects of phase shifts, differential cross sections, entanglement powers, and EOSs of symmetric nuclear matter and pure neutron matter. 

It was found that the phase shifts from chiral potentials have obvious differences in contrast to the partial wave analysis from the experimental data and other meson-exchange potentials for the high partial wave channels at high input energy region. This is because the number of low energy constants in chiral potentials is limited by the perturbation order.  Low-energy constants at order $Q^4$ are just determined by the $D$-state channel. Therefore, the phase shifts at the $F$ channel are the predicted values. Furthermore, the regularized cut-off in the chiral potentials also has a non-negligible effect. In addition, the mixing angles between $^3S_1$ and $^3D_1$ channels from various $NN$ potentials are also quite distinguished, which are not observable in $NN$ scattering.

The differential cross sections from these potentials for the $pp$ scattering below $E_{lab}=100$ MeV are consistent with each other and produce the experimental data perfectly at the scattering angle from $0$ to $\pi$. Whereas, there are obvious differences around $\theta_{cm}=0$ and $\theta_{cm}=\pi$ once $E_{lab}>200$ MeV.  Especially, the descriptions of chiral potentials on observables become worse, which should be related to poor phase shifts at high partial waves. All of these potentials produced a similar entanglement power for $np$ scattering at $^1S_0$ and $^3S_1$ channels. Their spins flip around input momentum $p=14$ MeV. At very high momentum region, the entanglement power from $\mathrm{N^{4}LO}$ 550(EMN) potential is lower than others.

The equations of state of infinite nuclear matter from these potentials have similar behaviors below the nuclear saturation density in the framework of the BHF method. As the density increases, the differences among high-precision nuclear interactions become obvious. The chiral potentials with momentum regulators are very soft so that there are no saturation points for symmetric nuclear matter before nuclear density $\rho<0.5$ fm$^{-3}$. However, the saturation properties from other potentials are far from the empirical data, indicating that the extra repulsive contributions from the three-body force are required.

Although the high-precision $NN$ potentials were claimed to describe the scattering observables with an accuracy of $\chi^2/N\sim 1$, they are still quite distinct in high-energy and high-density regions. It is necessary to reduce their uncertainties in these extreme conditions so that we can apply them to the investigations on the nuclear many-body system.

\section{Acknowledgments}
This work was supported in part by the National Natural Science Foundation of China	 No. 12175109, the Natural Science Foundation of Tianjin (Grant  No: 19JCYBJC30800), and the Natural Science Foundation of Guangdong Province (Grant  No: 2024A1515010911).

\end{document}